# Collimation method studies for next-generation hadron colliders

Jian-Quan Yang, [1,2] Ye Zou, [3] and Jing-Yu Tang[1,4*]

[1]*Institute of High Energy Physics, CAS, Yuquan Road 19B, Beijing 100049, People's Republic of China*
[2]*University of Chinese Academy of Sciences, CAS, Yuquan Road 19A, Beijing 100049, People's Republic of China*
[3]*Uppsala University, Uppsala 75120, Sweden*
[4] *Dongguan Neutron Science Center, Dongguan 523803, People's Republic of China*

Abstract: In order to handle extremely-high stored energy in future proton-proton colliders, an extremely high-efficiency collimation system is required for safe operation. At LHC, the major limiting locations in terms of particle losses on superconducting (SC) magnets are the dispersion suppressors (DS) downstream of the transverse collimation insertion. These losses are due to the protons experiencing single diffractive interactions in the primary collimators. How to solve this problem is very important for future proton-proton colliders, such as the FCC-hh and SPPC. In this article, a novel method is proposed, which arranges both the transverse and momentum collimation in the same long straight section. In this way, the momentum collimation system can clean those particles related to the single diffractive effect. The effectiveness of the method has been confirmed by multi-particle simulations. In addition, SC quadrupoles with special designs such as enlarged aperture and good shielding are adopted to enhance the phase advance in the transverse collimation section, so that tertiary collimators can be arranged to clean off the tertiary halo which emerges from the secondary collimators and improve the collimation efficiency. With one more collimation stage in the transverse collimation, the beam losses in both the momentum collimation section and the experimental regions can be largely reduced. Multi-particle simulation results with the MERLIN code confirm the effectiveness of the collimation method. At last, we provide a protection scheme of the SC magnets in the collimation section. The FLUKA simulations show that by adding some special protective collimators in front of the magnets, the maximum power deposition in the SC coils is reduced dramatically, which is proven to be valid for protecting the SC magnets from quenching.

## I. INTRODUCTION

For high energy proton-proton colliders, SC magnets are essential to achieve the magnetic strength required to reach higher center of mass energy. These magnets have an increasing sensibility to particle losses, which scales in level with magnetic strength. A tiny fractional loss of full beam in SC magnet coils, even the radiation from the particles lost in other locations, could lead to a quench, thus any significant beam loss in the SC magnets (also called cold magnets with respect to warm magnets for room-temperature magnets) must be avoided. Large beam losses could also cause serious damage to other accelerator components. However, beam losses cannot be completely suppressed because of various beam dynamic processes, such as beam-beam interactions, transverse and longitudinal diffusion, residual gas scattering and so on [1]. Therefore, besides strictly controlling beam losses and very reliable beam abort system, a robust and extremely-efficient collimation system is necessary to safely dispose of beam losses. At LHC [2], the highest energy collider in the world nowadays, a very complex collimation system with the multi-stage collimation method was designed, and a total of 132 pure collimators were designed, including primary collimators, secondary collimators, absorbers, tertiary collimators and other protection collimators, which have been and will be installed in a phased approach [3-5]. During the LHC Run I, the beam energy was up to 4 TeV and stored beam energy was up to 143 MJ, the collimation system with so-called tight settings [6] accomplished its tasks very well [7-8], a record cleaning inefficiency below a few $10^{-4}$ was achieved in the cold regions where they were filled with SC magnets and the strictest control of beam losses was required. As to the LHC Run II, the beam energy was increased from 4 TeV to 6.5 TeV and a shorter bunch spacing of 25 ns was adopted. With higher stored energy of about 270 MJ, the performance of the collimation system is still very good [9-10]. However, when LHC is upgraded to HL-LHC with higher stored energy, more particle losses in the DS downstream of the transverse collimation insertion is considered threatening to the local cold dipoles [8, 11]. These losses are due to the protons experiencing single diffractive interactions in the primary collimators [12]. Such protons can survive the interactions and emerge from the collimator jaws with their momentum modified only slightly in direction, but significantly in magnitude. In other words, this process converts the transverse halo particles into off-momentum halo particles. Thus, those protons will be able to escape from the transverse collimation system and are lost as soon as they reach the downstream DS section. In order to largely reduce irradiation dose rate to the SC magnets at the downstream DS section of the betatron collimation insertion (IR7) for each beam, two local collimators need to be added in the DS section. However, there is not enough space for additional collimators due to the compact design in the DS region. One solution to create space for additional collimators is to movethe cold magnets in the DS section [13]. Another solution is to replace the two original main dipoles of 8.3 T by two new shorter dipoles with new $Nb_3Sn$ magnet technology which can work at 11 T [14].

___________________________
[*]Corresponding author.
tangjy@ihep.ac.cn

For the design of future proton-proton colliders, due to the increasing probability of single diffractive interaction with the increase in energy [15-18], the problem of beam losses in the DS where the dispersion starts to increase becomes more important and should be treated with greater care. For FCC-hh, an analogous solution to the HL-LHC with local dedicated protection collimators in the DS [19], and the problem of DS losses can be almost solved. However, it would be costly to make this kind of arrangement in the DS regions, since the same space arrangement will be applied to all the arcs due to the symmetry of the ring. In this paper, a different approach is presented, which arranges both the transverse and momentum collimation in the same cleaning insertion. In this way, the downstream momentum collimation system will clean off the particles with large momentum deviation including those experience single diffractive interactions with the primary transverse collimators. In this way, one can get rid of beam losses in the DS regions and design the arc lattice as compact as possible. However, the challenge of this method is how to join two different collimation sections. In general, the transverse collimation section is designed to be approximately dispersion-free, but relatively large dispersions are required at the locations of primary collimators in the momentum collimation section [20]. Different from the momentum collimation section at the LHC where dispersion is intentionally designed non-zero between the two adjacent DS sections, thus a chicane-like and achromatic design for the momentum collimation section is adopted here. The schematic of the method is depicted in Fig. 1, where arc dipole magnets with a simplified design of only single aperture instead of twin aperture are used, where the associated cryomodules should be designed specially to allow the pass-through of another beam pipe. In this way, sufficient longitudinal space can be available here to add necessary protective collimators and shielding in room-temperature. Detailed studies including lattice design and multi-particle simulations using the MERLIN code [18, 21-23] have been carried out to check the validity of the method. The result shows that this method works as expected and the beam losses at the downstream DS can be suppressed. In addition, SC quadrupole magnets with special protection are being considered to provide more phase advance in the transverse collimation section, where room-temperature magnets are used due to high radiation dose rate. This measure can enhance the transverse collimation efficiency.

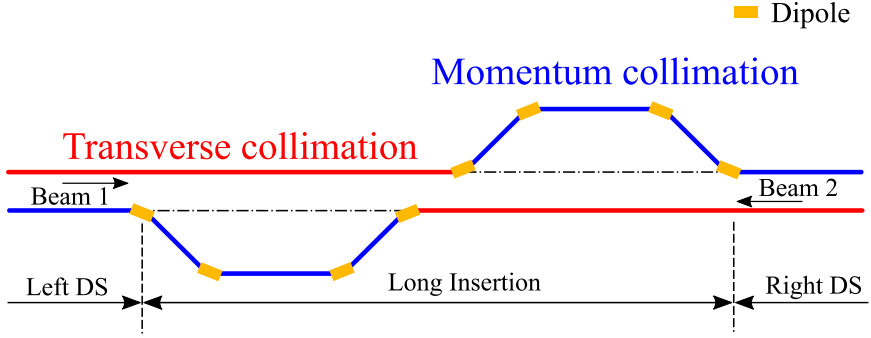

FIG. 1.    Layout of the combined transverse and momentum collimation method

For next-generation hadron collider, the above method is a general and applicable solution for collimation system. As the first conceptual approach phase, the studies presented here are mainly based on the parameters of the Super Proton-Proton Collider (SPPC), which is the second phase of the CEPC-SPPC project [24]. The layout and main parameters of the SPPC are given in the Appendix.

## II. LATTICE FOR THE COMBINED BETATRON AND MOMENTUM COLLIMATION

The top beam kinetic energy of SPPC in the baseline design is 37.5 TeV, which is about five times of that at LHC. The energy of halo particles is too high to be dissipated in a straightforward way. One general method to stop high-energy protons is to use multi-stage collimators. Depending on the collimators' functions, they are divided into several families. The primary collimators will intercept or scatter the primary halo particles, and the secondary collimators will intercept the secondary beam halos that are formed by the particle's interaction with the primary collimators. Sometime tertiary collimators will intercept the so-called tertiary beam halos (i.e. what emerges from the secondary collimators). At LHC, tertiary collimators are placed at the IPs, which define the minimum machine aperture in the inner triplet magnets of the IR region and protect the bottlenecks, represented by the inner triplets at the interaction points with the smallest $\beta^*$ (the beta function at the interaction point). The absorbers will stop the hadron showers from the upstream collimators and additional collimators are used to protect the SC magnets.

As critical components in ensuring the safe operation of an accelerator like SPPC, based on the experience at LHC [19], the material of collimators would have to meet requirements: good conductivity to reduce coupling impedance, high robustness to resist abnormal beam impacts, good absorption ability for cleaning efficiency [4]. Unfortunately, not all the three conditions can be fulfilled by the same material. A robust material, such as graphite, would increase the coupling impedance, which is important for collective beam instabilities and limits the machine performance. On the other hand, a good conductor, such as copper, is not robust enough, which means that the collimator jaws can be damaged even in the normal operation mode. Thus different materials will be used. As the closest objects to the circulating beam, primary and secondary collimators must withstand the highest dose of deposited energy without permanent damage. For this reason, they are made of robust carbon fiber-carbon composite. However, referring the collimation study for FCC-hh [25, 26], for an extremely high stored beam energy of about 10 GJ, assuming the total beam loss within 0.2 hour in the transverse collimation section, primary or secondary collimators have to resist power load of several hundreds of kW, it is extremely challenging for the robustness of the collimators. As for the tertiary collimators and absorbers, due to lower heat power load, they are made of high Z material, such as copper and tungsten, which can absorb particles efficiently and reduce the impedance relatively. Meanwhile, the number of collimators and sharing of phase space coverage ensure that the large level of energy deposition is distributed among them, avoiding single device overloaded. Thus, the location for each collimator needs to be optimized according to the $\beta$ functions, in order to obtain larger gap openings to reduce impedance issues and obtain appropriate phase advance between collimators to improve the cleaning efficiency.

### A. Requirements for the lattice design for transverse collimation

At LHC, collimators represent more than 90% of the impedance of all the accelerator components [27], and they produce the transverse wall impedance which scales inversely proportional to the third power of collimator gap size. Thus one effective method to reduce the impedance is to enlarge the collimator gap, which means that the collimators must be located at large $\beta$ values in the case of the unchanged ratio of gaps over beam size in $\sigma$ (normalized transverse rms beam radius). In addition, with a larger $\beta$, the same change in Courant-Snyder invariant means a larger change in amplitude, which enhances the impact parameter and reduces the out-scattering probability. Therefore, the $\beta$ function is required to be larger in the collimation insertion than in the arcs. To have high collimation efficiency in a multi-stage collimation system, the phase advance between different stages of collimators is also very important, thus a long insertion is needed to produce enough phase advances.

For proton accelerators, transverse collimation plays a major role relative to momentum collimation, thus the former has higher requirements for the lattice design and collimators and will withstand higher radiation doses. According to the principles of two stage betatron collimation [28], in one-dimensional case, the optimal phase advance $\mu_{\text{opt}}$ should satisfy

$$\cos\mu_{opt} = \pm n_1 / n_2 , \quad (1)$$

where $n_1$ and $n_2$ donate the apertures of primary and secondary collimators in unit of $\sigma$, respectively. For two-dimensional case, it becomes complicated. At LHC, the long straight sections offer a phase advance $\Delta\mu_{x,y} \approx 2\pi$. In order to minimize the maximum betatron amplitudes of protons surviving the collimation system, the longitudinal positions of collimators (same as the phase advance between collimators) were optimized with the code DJ [29, 30].

For next-generation colliders, a reasonable idea to improve transverse collimation efficiency is adding one more collimation stage to the four-stage collimation system used at LHC, which means larger phase advance needed in the transverse collimation section. On the premise of guaranteeing the beta functions without significantly increasing the total length of the collimation section, replacing warm quadrupoles by cold quadrupoles in the section is the only viable method. Next, we will explore the feasibility of this method in details, together with the design scheme using conventional warm quadrupole magnets.

## B. Requirements for the lattice design for momentum collimation

In general, a particle reaches the primary collimator with a mixing of betatron amplitude and momentum deviation. So we can define the largest momentum deviation $\delta_{\max}$ with which a particle can pass through the primary momentum collimator by the following formula [28]

$$\delta_{\max} = \frac{n_1 \sqrt{\varepsilon}}{\eta_1}, \tag{2}$$

where $n_1$ denotes the aperture of the primary momentum collimator in unit of $\sigma$ (containing the dispersive part), $\varepsilon$ denotes the geometric emittance in rms while $\eta_1$ denotes the normalized dispersion at the collimator. If the maximum normalized dispersion in the primary momentum collimation section is larger than the one at the DS or the whole arc section, in principle there will be very little beam losses in the downstream DS section or even all the arc sections, based on the fact that the arc aperture is larger than $n_1$. For more specific considerations, the normalized dispersion at the primary momentum collimator must satisfy [31]

$$\left|\eta_{D,\mathrm{prim}}(n_1)\right| \geq \frac{n_1 \eta_{D,\mathrm{arc}}}{A_{\mathrm{arc,inj}}(\delta_p = 0) - \left(n_2^2 - n_1^2\right)^{1/2}}, \tag{3}$$

to avoid cold losses at the DS or in the arc, where $A_{\mathrm{arc,inj}}(\delta_p = 0)$ denotes the arc aperture for on-momentum particles in unit of $\sigma$; $\eta_{D,\mathrm{arc}}$ is the normalized dispersion with errors in the focusing quadrupole magnets; $n_1$ and $n_2$ denote the apertures of primary and secondary momentum collimators. In addition, for ensuring that the cut of the secondary halo is independent of the particle momentum, the dispersion derivative $\eta'_x$ at the position of the primary momentum collimator must satisfy [28]

$$\eta'_x = 0. \tag{4}$$

As the momentum collimation deals with much smaller halo particles than the transverse collimation does and the impact parameters at the primary momentum collimators are also much larger, the collimation efficiency is not a problem.

## C. Lattice Scheme I with room-temperature quadrupoles in the transverse collimation section

In order to confirm the effectiveness of the novel method, the SPPC collimation system is used as a test-bench. As shown in Appendix, one can see that two very long straight insertions, LSS1 and LSS5, with a length of 4.3 km are used for collimation and extraction, respectively. In a dedicated collimation section, warm quadrupoles are usually used for their high radiation resistance. However, for very high energy proton beams, the focusing strength is a problem. Thus we use quadrupole groups here each representing several quadrupole units arranged together and acting as one quadrupole. For the momentum collimation section, in order to produce the required dispersion, four groups of cold dipoles of arc dipole type are used. Meanwhile, cold quadrupoles are also used to control the betatron functions in the limited space. Figure 2 shows the optics in Lattice Scheme I for the SPPC collimation system, which is similar to the lattice design in FCC-hh to some extent [19, 32]. The main parameters are listed in Table I.

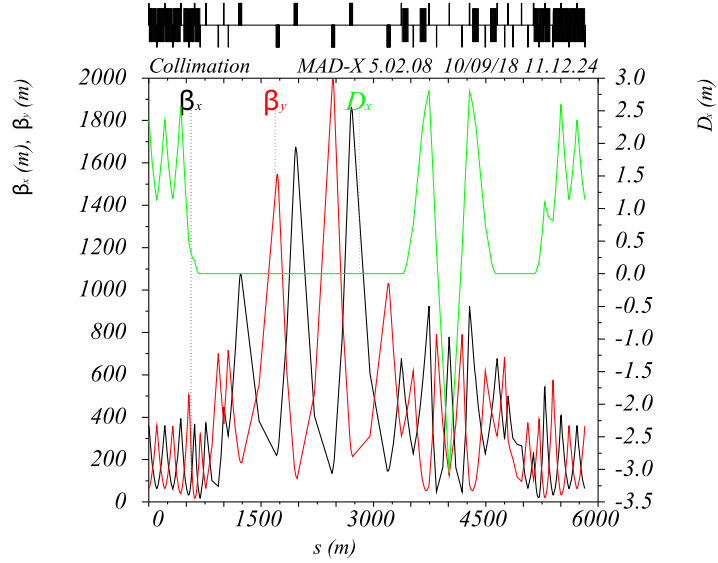

FIG. 2.   The betatron and dispersive functions of the collimation insertion in Lattice Scheme I

TABLE I.   Basic parameters of the collimation insertion in Lattice Scheme I

| Parameter | Unit | Betatron collimation | Momentum collimation |
|---|---|---|---|
| Insertion length | km | 2.55 | 1.75 |
| Maximum beta function $\beta_x/\beta_y$ | km | 1.86/1.99 | 0.92/0.79 |
| Phase advance $\mu_x/\mu_y$ | rad | $1.89\pi/1.91\pi$ | $2.27\pi/2.14\pi$ |
| Quadrupole length | m | 3.3 | 6.0 |
| Maximum quadrupole strength | $\times 10^{-3}$ m$^{-2}$ | 0.19 | 2.2 |
| Quadrupole aperture | mm | 50 | 56-80 |
| Maximum quadrupole magnetic field | T | 0.6 | 7.8 |
| Dipole length | m | - | 14.45 |
| Dipole magnetic field | T | - | 12 |
| Dipole aperture | mm | - | 50 |
| Maximum normalized dispersion | m$^{1/2}$ | - | -0.124 |

### D. Lattice Scheme II with only SC magnets

For a multi-stage collimation system, the primary and secondary collimators generate secondary and tertiary halo particles that extend several $\sigma$ beyond the collimator settings, and some of them escape from the collimation insertion and are lost on the inner SC triplets at IPs where the apertures are reduced by the very large $\beta$-functions. At LHC, in order to locally provide additional protection from the tertiary halo [33], 16 tertiary collimators in pairs are installed at each side of the four experiment insertions. These tertiary collimators also can protect the triplets from the mis-kicked beams, for example, due to failures of the normal conducting separation magnets [34]. One source of the machine-induced backgrounds at the detectors is due to the upstream interaction of beam protons with residual gas molecules or collimators. According to the study at LHC [35, 36], the beam-gas interaction is the main contribution of background, higher than the beam-halo by one order of magnitude.

For SPPC, the stored energy in the beam is as high as 9.1 GJ per beam, about 25 times of that of the LHC at design energy, and the development of hadronic and electromagnetic shower become more intense due to higher proton energy. It is foreseeable that the tertiary halo in the machine will be much more severe. One more stage of collimators installed in the transverse collimation section will convert the tertiary beam halo into quaternary beam halo, thus can help to dilute the halo particles in the experiments and reduce the risk of quenching in SC inner triplets, and the experimental background level may be reduced more or less. However, when warm quadrupoles are used, there is not enough phase advance to add additional collimators due to the weak focusing strength, or significant space will have to be added. As the space is so precious, therefore, we try to apply SC quadrupoles in the transverse collimation section to create more focusing cells. These quadrupoles are very different from those in the arcs, they will be designed with enlarged apertures and lower pole strength (no higher than 8 T), and are somewhat comparable to the triplet quadrupoles used in the experiment insertions at LHC. In this way, much higher transverse collimation efficiency can be obtained, so that the probability of particle losses in the downstream momentum collimation section and the residual halo at the experiments will be reduced largely. The phase advance between the secondary and tertiary collimators should be similar as the one between the primary and secondary collimators, assuming that most of the tertiary beam halo particles are emitted from the secondary collimators. Figure 3 shows the lattice functions. The main parameters are listed in Table II.

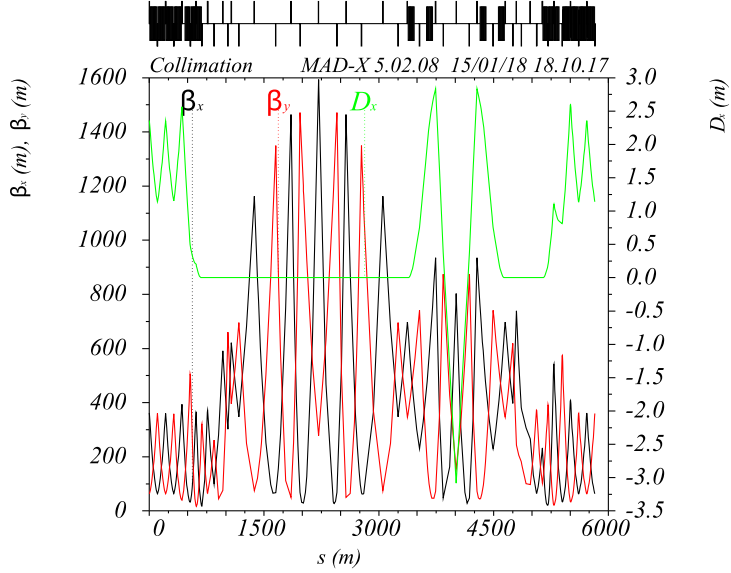

FIG. 3.　　The betatron and dispersive functions in the collimation insertion in Lattice Scheme II

TABLE II. Basic parameters of the collimation insertion in Lattice Scheme II

| Parameter | Unit | Betatron collimation | Momentum collimation |
| --- | --- | --- | --- |
| Section length | km | 2.55 | 1.75 |
| Maximum beta function $\beta_x/\beta_y$ | km | 1.59/1.47 | 0.93/0.87 |
| Phase advance $\mu_x/\mu_y$ | rad | 3.54π/3.31π | 2.27π/2.14π |
| Quadrupole length | m | 6.0 | 6.0 |
| Maximum quadrupole strength | ×10$^{-3}$ m$^{-2}$ | 1.8 | 2.2 |
| Quadrupole aperture | mm | 70-80 | 56-80 |
| Maximum quadrupole magnetic field | T | 8.0 | 7.8 |
| Dipole length | m | - | 14.45 |
| Dipole magnetic field | T | - | 12 |
| Dipole aperture | mm | - | 50 |
| Maximum normalized dispersion | m$^{1/2}$ | - | -0.124 |

Same as for Lattice Scheme I, we also need to consider the two collimation systems for each beam in one insertion. The distance between two beams is set to about 30 cm at the arcs, which is considered to be enough to install one collimator for one beam but cannot accommodate an additional collimator at the same location for another beam. In the momentum collimation section, the horizontal separation from the other beam is enlarged to 1.64 m which will allow the installation of the collimators for the two beams. Meanwhile, we apply SC quadrupoles with twin apertures for the two beams in the overlapping region with nominal separation. The layout of the collimation section is shown in Fig. 4.

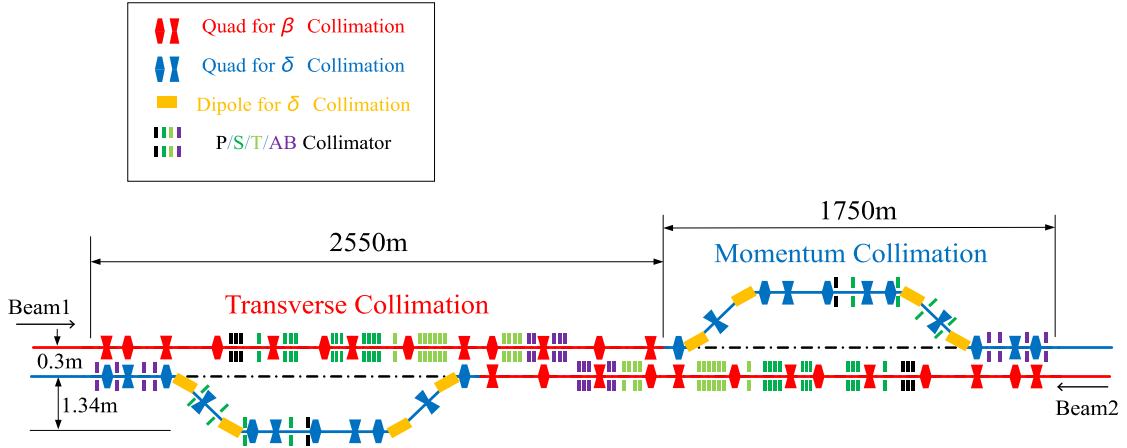

FIG. 4. Layout of the collimation insertion. P/S/T/AB denote primary collimator, secondary collimator, tertiary collimator and absorber.

## III. SIMULATION RESULTS OF THE SPPC COLLIMATION SYSTEM

### A. Collimation inefficiency

To quantify the performance of the collimation system more precisely, the local cleaning inefficiency $\tilde{\eta}_c$ which is the ratio of number $N_i$ of lost protons at any location of the ring in a given bin of length $L_i$ (set to 10 cm in general) over the total number $N_{tot}$ of lost protons [4]

$$\tilde{\eta}_c = \frac{N_i}{N_{tot} \cdot L_i}. \tag{5}$$

For slow and continuous losses, the circulating intensity in the machine can be described as a function of time $t$

$$N(t) = N_0 \exp\left(-\frac{t}{\tau}\right), \tag{6}$$

where $N_0$ is the nominal intensity, $\tau$ is the finite beam lifetime. At LHC, in order to ensure commissioning and performance in nominal running, conservative minimum lifetimes $\tau_{\min}$ are assumed as 0.2 hour at top energy and 0.1 hour at injection energy [37]. For an operation with the minimum beam lifetime $\tau_{\min}$, the total intensity $N_{tot}^q$ is limited by the quench limit $R_q$

$$N_{tot}^q = \frac{\tau_{\min} \cdot R_q}{\tilde{\eta}_c}. \tag{7}$$

where $\tilde{\eta}_c$ is the local cleaning inefficiency as defined in Eq.(5), the quench limit $R_q$ in unit of protons/m/s is related to the transmission capability and the maximum deposited energy density, which defines the allowed maximum local proton loss rates [38]. Figure 5 shows the maximum total intensity at the quench limit as a function of the local cleaning inefficiency, assuming that minimum beam lifetimes of 0.1 hour at the injection energy and 0.2 hour at the top energy must be satisfied just like at the LHC. In the baseline design of SPPC, the SC magnets in the arcs use the full iron-based HTS technology [39], and the field strength of the main dipoles is 12 T. However, as the magnet technology is still being developed [40], the quench limits is not yet available. In this article, the quench limit value $R_q$ for the SPPC arc magnets is estimated by the following formula [41]

$$R_q = 1.7 \cdot 10^8 E^{-\frac{3}{2}}, \tag{8}$$

where the energy $E$ is in unit of TeV. The same scaling was applied in the FCC-hh design [19] from the NbTi technology at LHC to the Nb$_3$Sn technology at FCC-hh, assuming the same quench level 5 mW/cm$^3$ [38, 42]. Thus the quench limit $R_q$ is estimated as $0.74 \times 10^6$ protons/m/s at the top energy 37.5 TeV and $0.56 \times 10^8$ protons/m/s at the injection energy 2.1 TeV. From Fig. 5, it is noted that the SPPC at top energy has the most stringent requirements of the cleaning inefficiency, where the vertical line in black is for the design goal of this collimation study, which means that the nominal intensity of $1.5 \times 10^{15}$ protons per beam at top energy requires a cleaning inefficiency of $3.55 \times 10^{-7}$ m$^{-1}$, that is more stringent than that at the LHC by about one order.

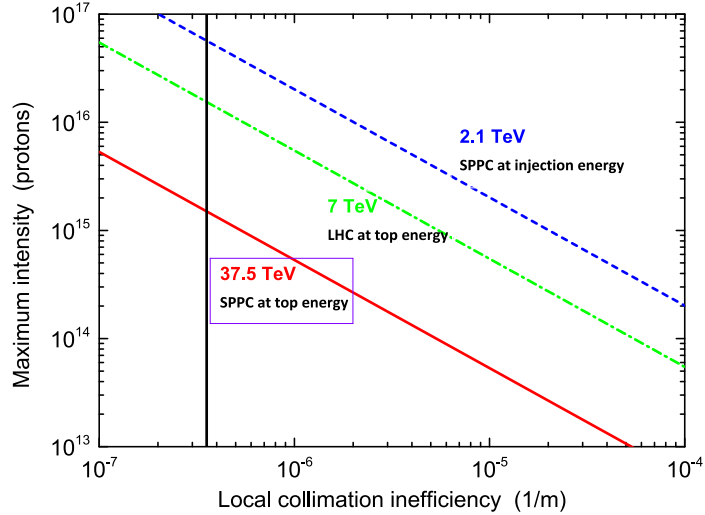

FIG. 5. The maximum total intensity is shown as a function of the local cleaning inefficiency for SPPC injection, top energy, and the LHC top energy.

### B. Simulation results of the beam loss distributions

Multi-particle simulations using the two lattice schemes described in Sections IIC and IID have been carried out with the code MERLIN [], which is a C++ accelerator library easily to extend and modify. Its organization structure can be found in References [43]. This code has a good agreement with the well-known collimation version SixTrack+K2 after the benchmarked work [44]. In the code, protons are considered lost if they undergo an inelastic interaction within the collimator jaws or if they intercept the mechanical beam pipe. The local cleaning inefficiency $\tilde{\eta}_c$ is used as the measure of the performance for collimation simulations.

Besides the arc sections, only the functional lattices for the collimation and experiment insertions have been used and all the other insertions, such as RF, injection and extraction insertions, are replaced by periodic FODO structures. The physical aperture in the arcs is set to be the inner aperture of the beam screen [45] that is used to absorb the synchrotron radiation, and its cross section is a superposition of an ellipse and a rectangle with a mean radius of about 15 mm. As for the transverse collimation section, the apertures of the warm quadrupoles are 60 mm or larger than 85 $\sigma$ in Lattice Scheme I, and the cold quadrupole apertures are enlarged to 70-80 mm or larger than 130 $\sigma$ in Lattice Scheme II, which are about the same as the triplet magnet apertures in the experiments regions. In the momentum collimation section, the quadrupole magnet aperture is enlarged slightly on the premise that the pole magnetic field does not exceed the preset value 8 T.

The collimator parameters in the simulations are shown in Table III, where T for transverse, M for momentum, P for primary, S for secondary, the second T for tertiary, Q for quaternary, AB for absorber, C for collimator, the collimator settings are quoted from the LHC design settings [34] and Run I operational settings [46]. The locations of the four-stage collimators in the collimation insertion are shown in Fig. 4. As the first approach for the transverse collimator settings, the physical gaps and phase advances are set as the same as the ones at LHC for just verifying the effectiveness of the collimation method, especially in cleaning particles related to the single diffractive effect. The primary momentum collimators are placed just downstream of the middle quadrupole between the second and third groups of dipoles to satisfy Eq.(3) and Eq.(4), where the normalized dispersion is close to the maximum. The other momentum collimators are placed further downstream with the same phase advances as used in the current LHC momentum collimation. 11 tertiary collimators have been added in Lattice Scheme II, but the jaw locations and orientations have not been optimized. In addition, two quaternary collimators are installed before the inner triplet magnets at each experiment region, which are to intercept the residual halo and to protect IR bottlenecks.

TABLE III. Collimator parameters for SPPC

| Collimator Acronym | Length (m) | Number | Aperture ($\sigma$) | Material | Lattice schemes |
|---|---|---|---|---|---|
| TPC | 0.6 | 3 | 6 | Carbon | I, II |
| TSC | 1.0 | 11 | 7 | Carbon | I, II |
| TTC | 1.0 | 11 | 8.3 | Copper | II |
| TAB | 1.0 | 5 | 10 | Tungsten | I, II |
| TQC | 1.0 | 4 | 10 | Tungsten | I, II |
| MPC | 0.6 | 1 | 12 | Carbon | I, II |
| MSC | 1.0 | 4 | 15.6 | Carbon | I, II |
| MAB | 1.0 | 4 | 17.6 | Tungsten | I, II |

In order to increase the accuracy of calculating the local cleaning inefficiency, 100 million protons are tracked for 300 turns in the SPPC ring, in which the initial beam distribution is represented by so-called halo distributions for saving the computing time. For example, for the horizontal halo collimation, the horizontal distribution is presented by two short arcs with a radius being the TPC half-gap, and the vertical distribution is a cut Gaussian at 3 $\sigma$, just as shown in Fig. 6. The impact parameter at the primary collimators is chosen as 1 μm, which is used for negligible emittance growth from the previous turn, referring to the setting in the simulation of the LHC collimation system [41]. Based on the above parameter settings, the simulations are carried out for both the horizontal and vertical halo collimations. This simulation method and assumptions can maintain a good computing performance, which have been illustrated in references [41, 46-47].

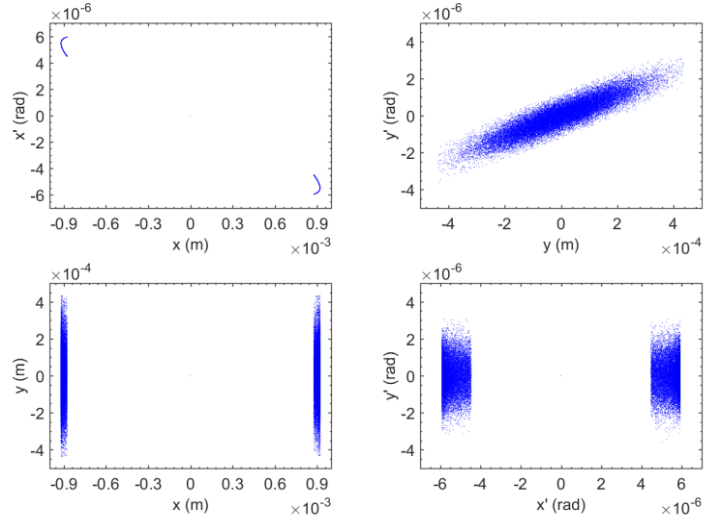

FIG. 6. Initial horizontal halo distribution for collimation simulations

Figures 7 and 8 show the proton loss distribution in the collimation insertion using Lattice Schemes I and II, with the initial horizontal and vertical halo distributions, respectively. One can see that there are still cold-area losses at the dipoles where the dispersion starts to rise, which can be foreseen due to the protons with large momentum deviations, related to the single diffractive effect. One can find the detailed comparison of proton losses in the first and second groups of cold dipoles with different lattice design schemes and initial halo distribution in Fig. 9.

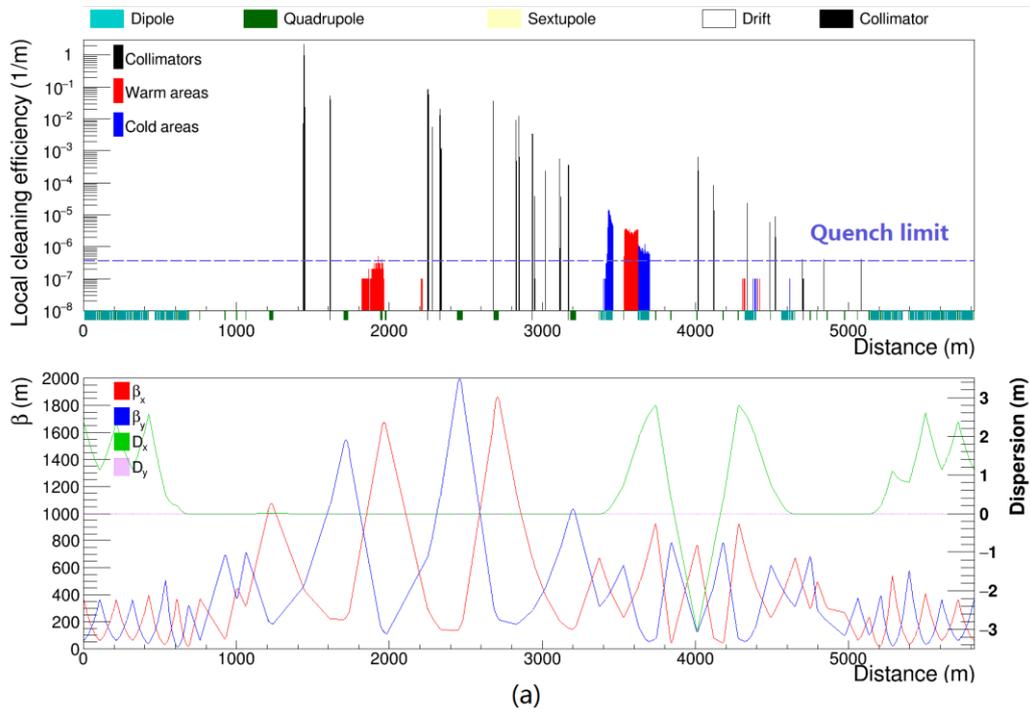

(a)

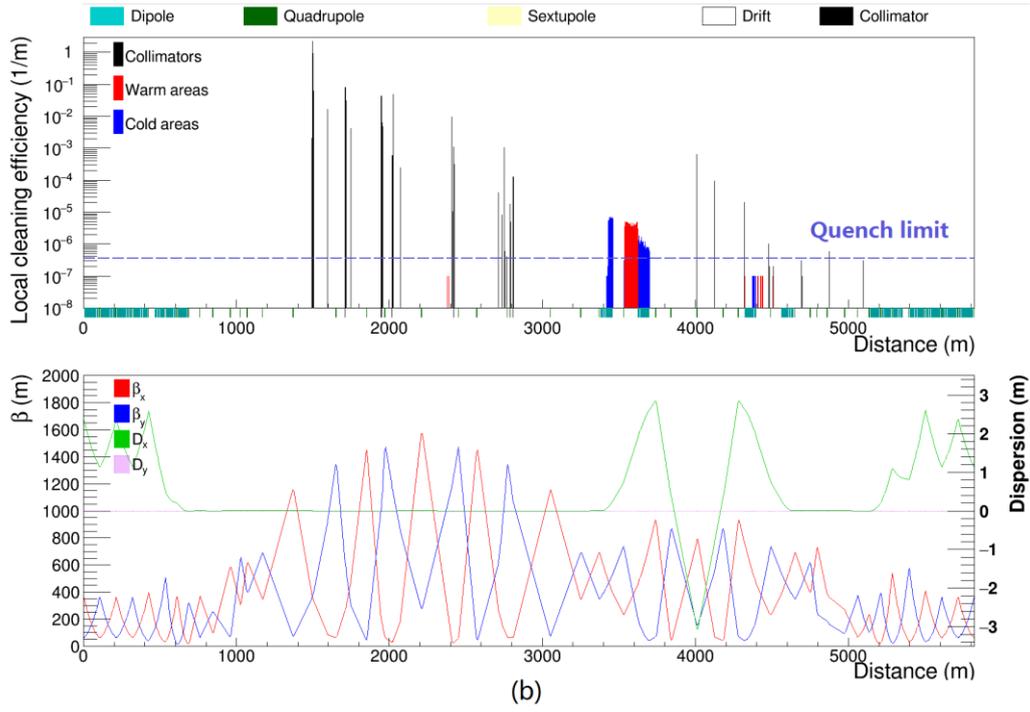

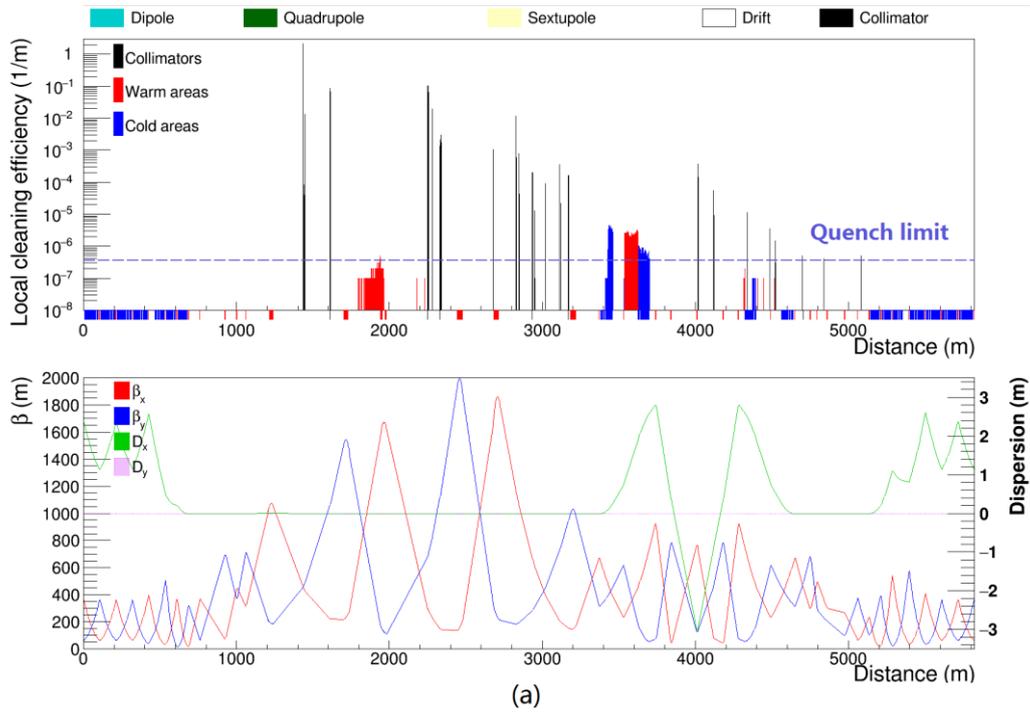

FIG. 7. Proton loss maps in the collimation insertion with an initial horizontal halo distribution, using Lattice Scheme I (a) and II (b)

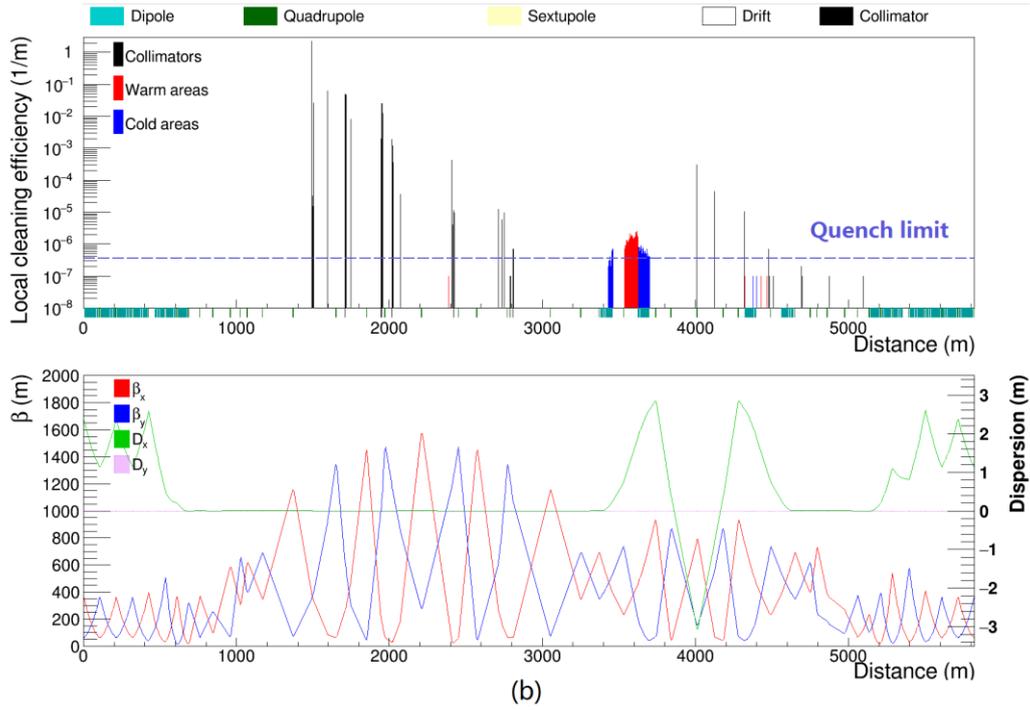

FIG. 8. Proton loss map in the collimation insertion with an initial vertical halo distribution, using Lattice Scheme I (a) and II (b)

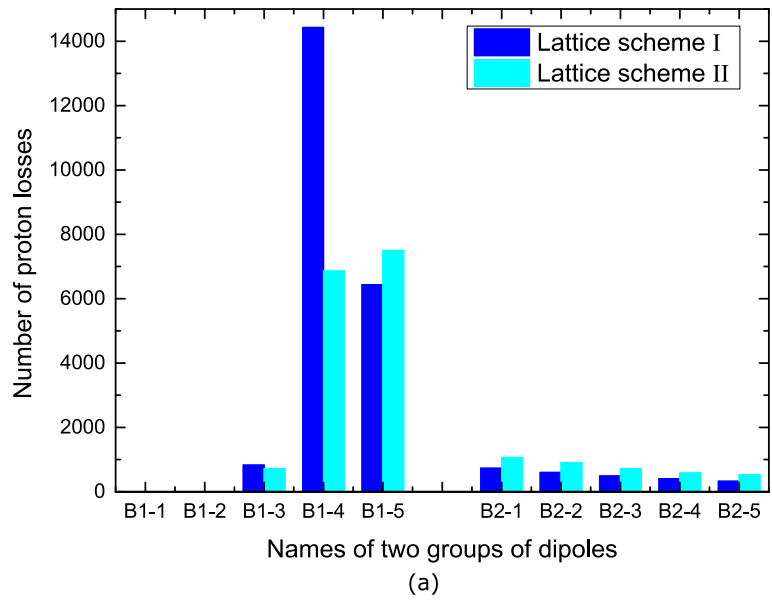

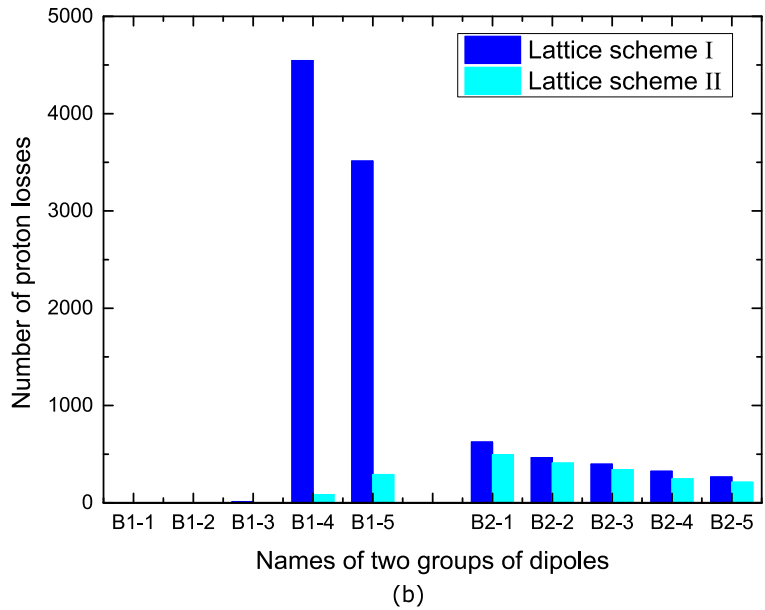

FIG. 9. Proton losses in the first and second groups of dipoles with two lattice schemes, with an initial horizontal (a) or vertical (b) halo distribution. The total particles in the simulations are 100 million.

With an initial horizontal halo distribution, the proton losses can be reduced to half by introducing 11 tertiary collimators, but it could still lead to cold dipole quenches if no further protection measures are made. In contrast, with initial vertical halo distribution, the tertiary collimators can reduce the proton losses by about one order of magnitude.

As shown in Fig. 7, for an initial horizontal halo distribution, one will see important proton losses at the cold dipoles due to the single diffractive effect, even with help of tertiary collimators. To solve the problem, some protective collimators (used as absorbers) can be placed here. Different from the arc DS regions where the lattice structure is very strict and the space is very tight, it is much easier to provide the space for the collimators in room temperature in the momentum collimation section. According to the positions of the lost particles, three protective collimators in Tungsten with an aperture of 10 $\sigma$ and length of 1 m, same as the one of the absorbers in the transverse collimation section, are placed there to intercept the particles related to the single diffractive effect. The specific locations are as follows: one protective collimator is placed between the third and fourth dipole magnets of the first dipole group to intercept particles with very large momentum deviation, and the cryostat for the dipole group is split into two parts to allow the insertion of the collimator in room temperature; another one is placed before the quadrupole between the first and second groups of dipoles to protect the quadrupole; the third one is placed in front of the second group of dipole magnets. Figure 10 shows the beam loss distribution in the collimation insertion with three protective collimators, with initial horizontal halo distribution, nearly all the proton losses in the cold regions disappear. Almost all the lost protons are in the collimators.

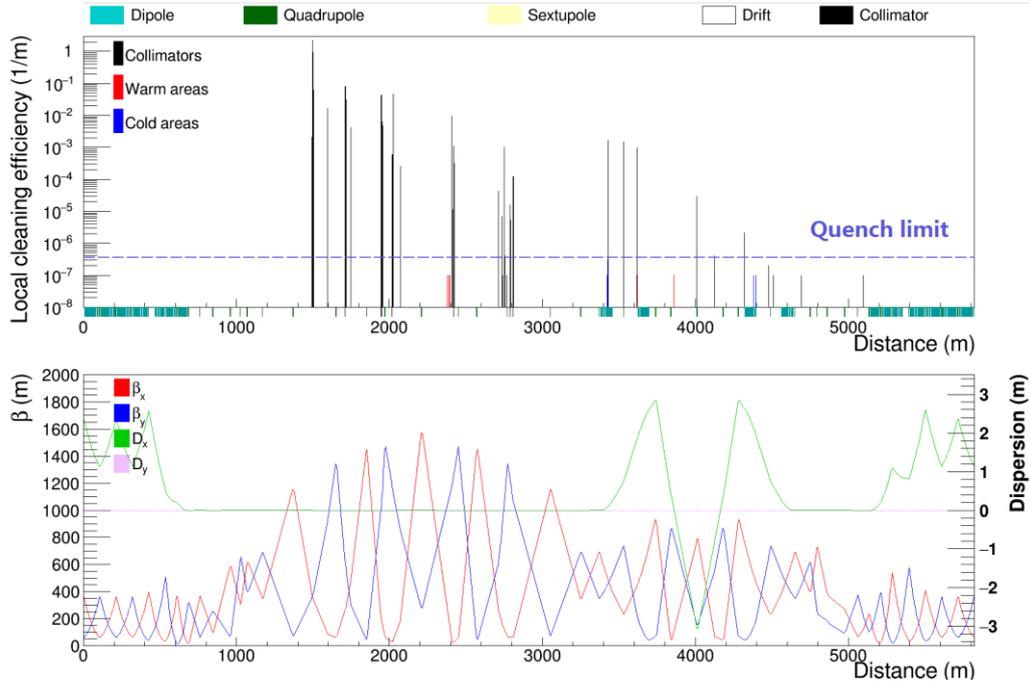

FIG.10. The proton loss distribution of the cleaning insertion with protective collimators

As mentioned earlier, the beam losses in the experiment regions are also a major concern. According to the simulation results, the tertiary collimators can intercept the tertiary halo effectively, as evidence the proton losses at the quaternary collimators are reduced by more than one order in the experiment region LSS7, and by four times in the experiment region LSS3, compared to the Lattice Scheme I; the results are shown in Fig. 11. This means it is much helpful to reduce the residual halo particles in the experiment regions by adding one more stage collimators in the transverse collimation section.

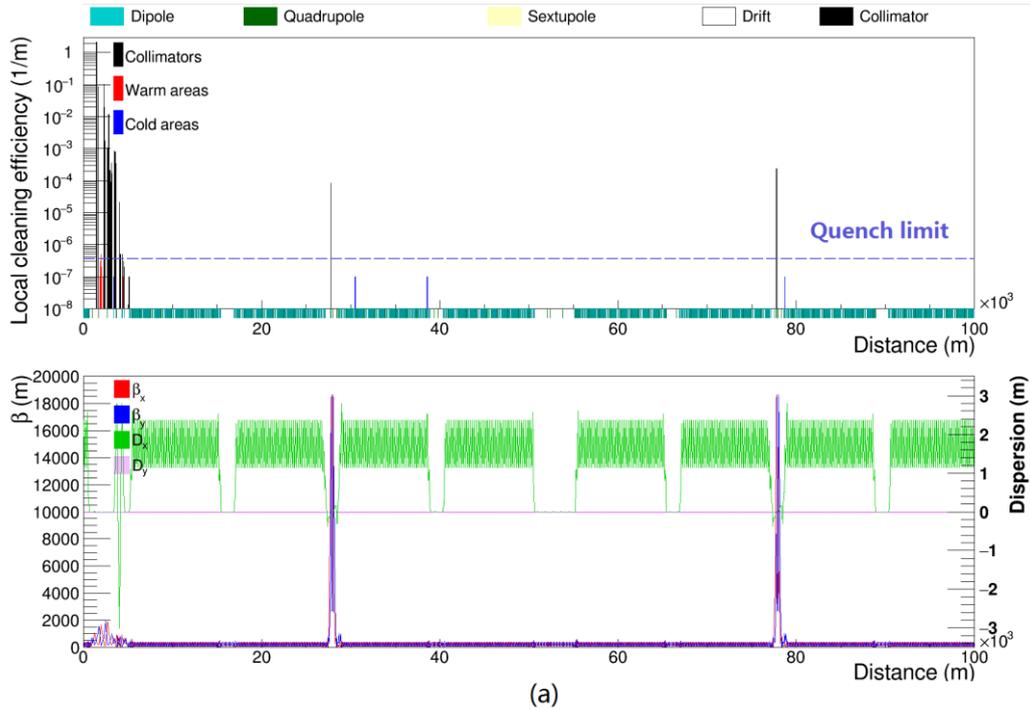

(a)

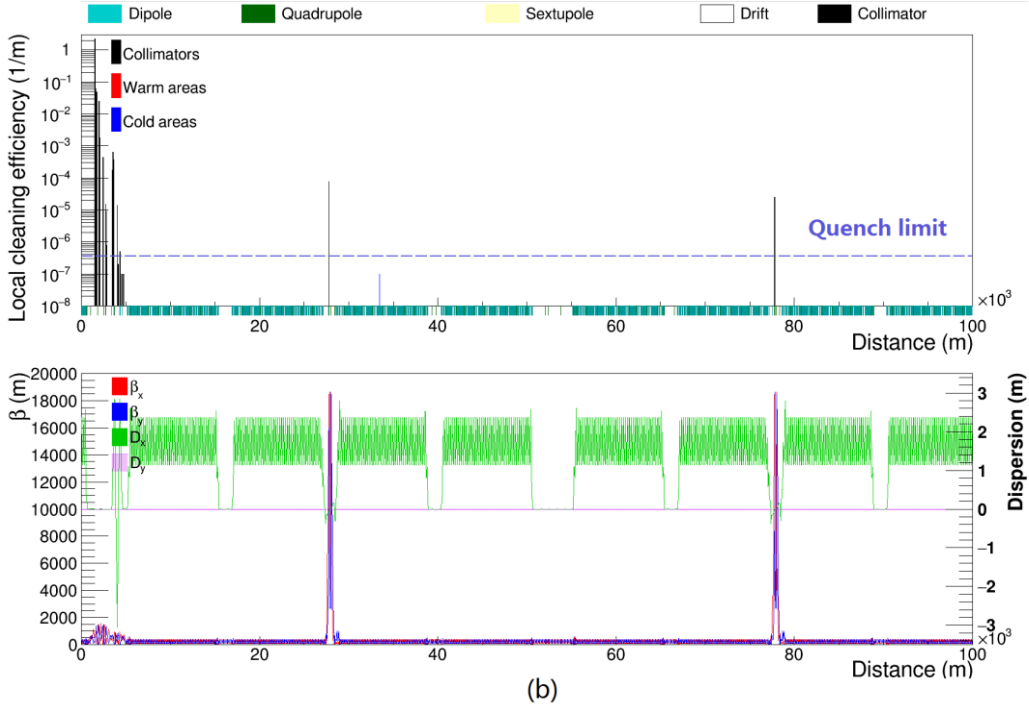

FIG. 11. Proton loss distribution along the full ring in lattice scheme I (a) and II (b), with initial vertical halo distribution

## IV. ANALYSIS OF THE ENERGY DEPOSITION IN THE SC MAGNETS OF THE COLLIMATION SECTION

### A. Quench limits

When a high-energy proton interacts with dense matter, the showering of the hadronic and electromagnetic cascades is the main process of energy deposition, which produces thousands of low-energy particles continuously until all of them are stopped in the matter and absorbed. These processes occur in the interactions between the primary protons and collimators or vacuum chamber. If these secondary showers deposit energy in the SC magnet coils then the local energy or power deposition may exceed the quench limit value, the SC magnets will experience a quench, from the SC state to the normal conducting state [48]. In general, quench limit is a function of local magnetic field, geometrical loss pattern, operating temperature, cooling conductions, and time distribution of beam losses [49]. In order to protect the SC magnets in the collimation section from quenches, it is very important to shield the particle showers and reduce the energy or power deposition in the magnet coils. In this section, we provide the protection schemes for the SC quadrupoles and dipoles which are used in the collimation section of Lattice scheme II. For simplicity, only steady state beam-loss is considered, the heat in the coils is constantly removed by the helium bath through the cable insulation [49].

To reduce the energy deposition in the SC coils, the cold quadrupoles in the transverse collimation section are designed with enlarged aperture and lower magnet field. On the one hand, the larger aperture means larger acceptance for the magnet to intercept as less as possible particles, on the other hand, the quench limit increases as the magnetic field decreases. As shown in Table II, the highest pole-tip magnetic field is 8 T, which is lower than the IR quadrupoles at LHC. Considering the Helium II and Helium boiling heat transfer mechanisms, which allow extracting more heat from the cable than the only solid conducting through the cable insulation, the estimated quench limits in the cable of cold quadrupoles ranges from 50-100 mW/cm$^3$ [49-50].

For the SC dipoles used in the momentum collimation section, the magnetic field is 12 T, which will use full iron-based HTS technology in the SPPC. However, some physical properties of the cable have yet to be determined so far. Thus, the conservative estimate of quench limits in the cable of cold dipoles is 5-10 mW/cm$^3$ [51], just same as the Nb$_3$Sn cable.

### B.  Energy deposition in the SC quadrupoles

The Monte Carlo analysis process for energy or power deposition includes the following steps.

1. The initial beam halo distribution is calculated with the MERLIN code, which records the coordinate information of the halo protons. This recorded data provide an input distribution for shower and energy deposition studies with the FLUKA code [52-53].

2. Particle-shower simulations with FLUKA are carried out to estimate the energy deposition in cold magnets for lost primary protons.

3. The power depositions in the most critical position of the coils for different beam loss scenarios are calculated. According to the tracking results with MERLIN, we know that 88% of protons are lost in the first turn, for the calculation results with FLUKA, the power deposition per bin is divided by 88%. The quench probability is evaluated based on the power deposition.

For power deposition evaluation in the SC quadrupoles in the transverse collimation section, observing the proton loss distribution as shown in Fig. 10, the first quadrupole downstream of the primary collimators is considered as the magnet which bears the greatest risk of quench. Thus the shower produced by the interaction between the halo and upstream collimators, including three primary collimators and one secondary collimator, is regarded as the main source of energy deposition in the cold coils of the quadrupole. The layout of a geometry model is shown in Fig. 12, where the input distribution is provided by MERLIN code with the initial horizontal halo distribution.

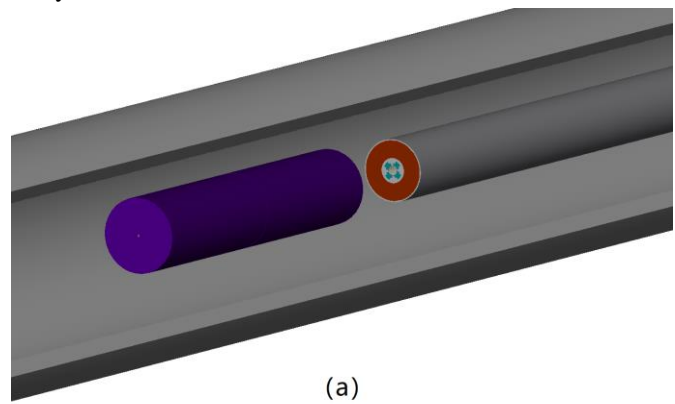

(a)

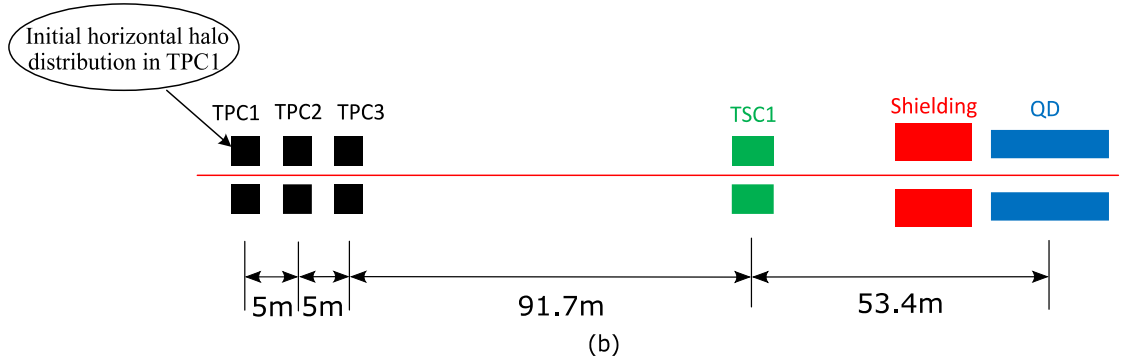

FIG. 12. Geometrical model of the shielding and defocusing quadrupole (a) and the positions of elements in FLUKA (b) for analysis of energy deposition in the first SC quadrupole after the primary transverse collimators, where the magnetic fields are not included in the simulation.

As mentioned in Section IIIB, in order to reduce the probability of particle losses in the SC coil of quadrupoles, they are designed with a wider aperture, and for the case of QD in Fig. 12 it is 80 mm. The material of coils is a mixture of 50% niobium-titanium and 50% copper. The cross section of the SC quadrupoles used in the transverse collimation section is shown in Fig.13, referring the insertion region wide aperture quadrupoles at LHC [48].

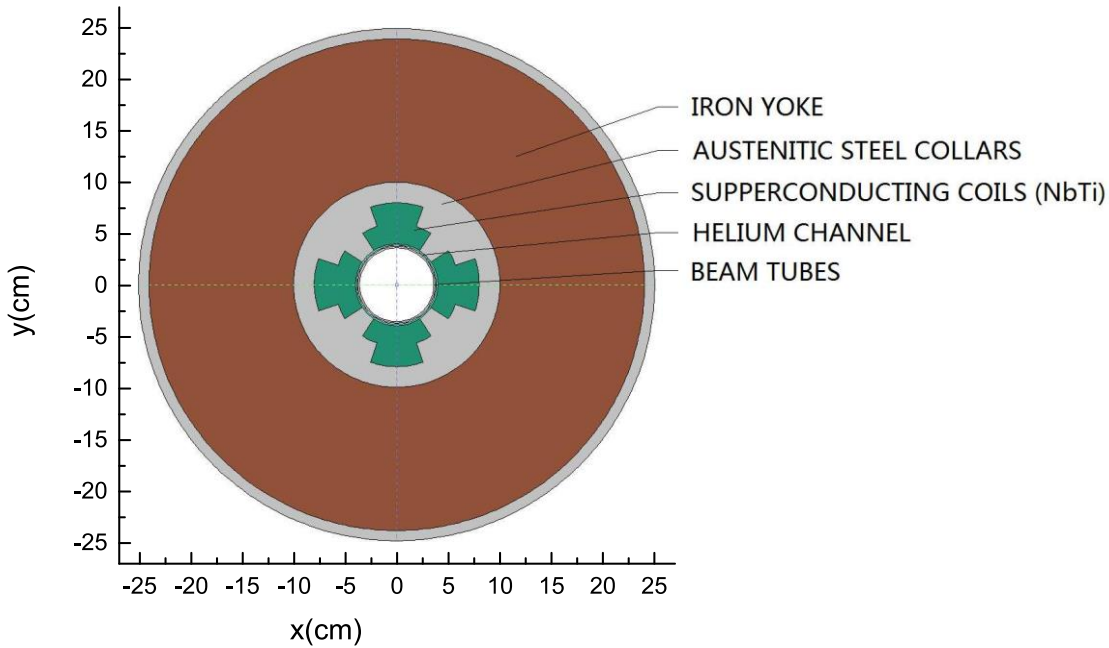

FIG. 13. Cross section of the SC quadrupole in transverse collimation section

As a high $Z$ material, Tungsten has been chosen as the shielding material, which can absorb the particle shower more effectively. In the geometry model, the shielding is placed at 1 m in front of the quadrupole, which is a hollow cylinder, with a length of 3 m and inner half-aperture of 10 mm or about 37 $\sigma$, outer radius is set to be 300 mm to cover the yoke of the SC quadrupole. This has proven to be tight enough to intercept the particle shower, but wide enough not to violate the multi-stage collimation hierarchy.

Referring to the energy deposition study for FCC-hh [26], with the assumption that the total beam is lost on collimation system within a time period of 0.2 hrs that is used for designing the LHC collimators, the maximum power on the dogleg warm dipoles is up to 1.1 MW, which is considered to be too high to cool the dipoles easily. This is the similar situation at SPPC. In this study, the loss rate of total beam power in one hour in the collimators is used to calculate the power deposition. This abnormal beam loss usually triggers the beam ejection mechanism into the external beam dump in a few seconds. Figure 14 shows the results of the maximum power deposition density, where the bins are chosen as a compromise between the calculation precision and time consuming: 0.5 cm in radius, 2° in azimuth and 5-10 cm in length (small bins for the region where the gradient is large)along the defocusing quadrupole QD. According to the calculation results by FLUKA, the total power on the shielding is up to 480 kW and the peak power density is about 1.3 kW/cm$^3$, which is located in the front face of the shielding. This value may be too high to bear for the tungsten material even for a short period, further optimization studies on the shielding material and structure, including cooling system for the shielding should be done in the future.

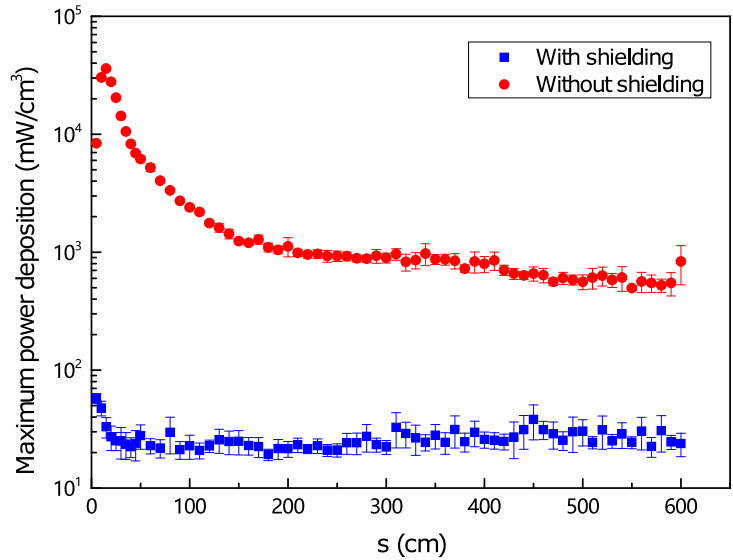

FIG. 14. Maximum power deposition density along the quadrupole QD after the primary collimators

Compared to the case without shielding (red line), the maximum power deposition along the QD with shielding is reduced by three orders of magnitude (blue line). In shielding cases, the peak power deposition is located at the first bin or the entrance part of QD, with a value of 57.7 mW/cm$^3$. The possible reason for the peak is that it comes from the shower emerged from the end part of the shielding block. An optimized method is to slightly increase the aperture of the rear half shielding, so-called step-like shielding. Figure 15 shows the simulation results after the optimization, where the aperture of the rear half shielding is enlarged from 10 mm to 10.5 mm for the step-like shielding. One can see that with step-like aperture the power deposition is reduced to below 30 mW/cm$^3$, which is safe from the quench limit value 50-100 mW/cm$^3$.

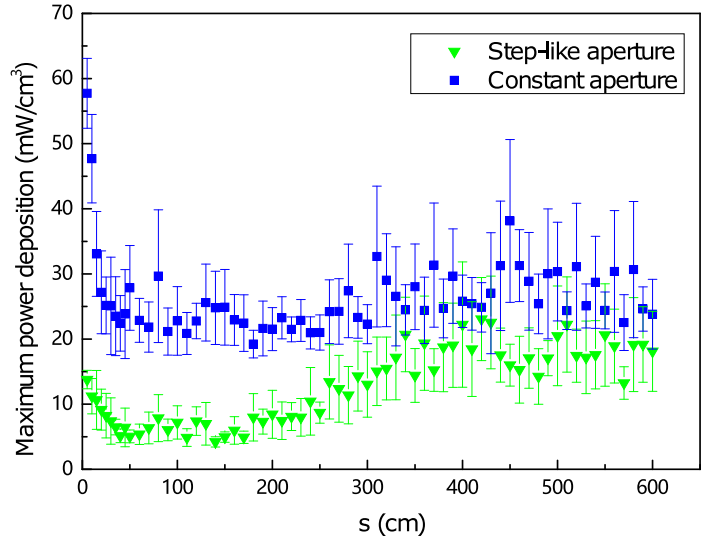

FIG. 15.  Maximum power deposition along the QD before and after the aperture optimization of the shielding

## C. Energy deposition in the SC dipoles

For the evaluation of quenching risk in the SC dipoles used in collimation system, the fourth dipole of the first group of dipoles, which is the closest to the first protective collimator, is considered the cold dipole which bears the highest dose of radiation. Figure 16 shows the 3D geometry model in FLUKA and the cross section of the SC dipoles, where the material of coils is chosen as the mixture of 50% Silver and 50% $SmAsFeO_{0.2}F_{0.8}$, which is one type of iron-based wires [54]. The input distribution of protons used in FLUKA is provided by the code MERLIN, which records the coordinate information of the protons lost in the first protective collimator. Here the upstream shower is not included, which is considered to be cleaned off by the transverse collimators and absorbers. For one hour beam lifetime, the power load on this protective collimator is 0.9 kW. Figure 17 shows the result of power deposition in the coils along the two most exposed dipoles, B1-4 and B1-5.

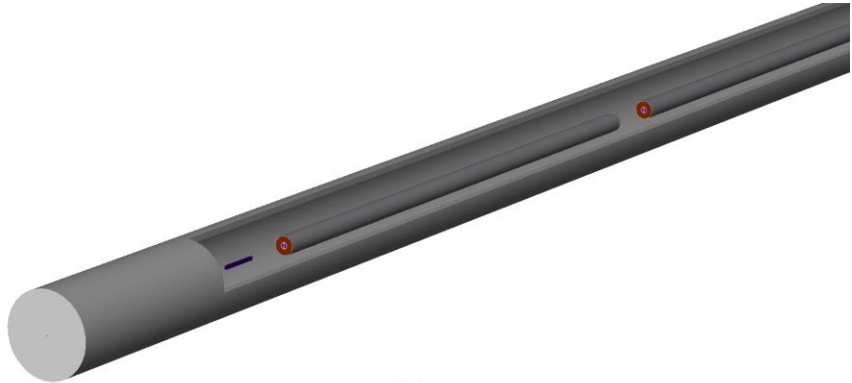

(a)

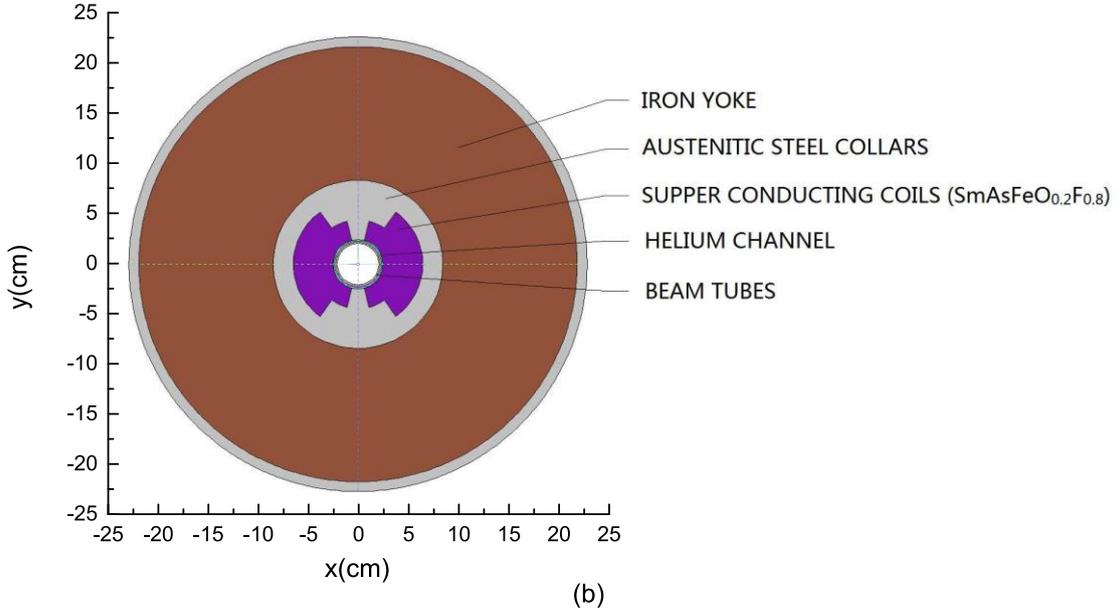

FIG. 16.  3D geometry model in FLUAK including the first protective collimator and two following SC dipoles B1-4 and B1-5 (a) and Cross-section of the SC dipoles in the momentum collimation section (b), where the magnetic fields are not included.

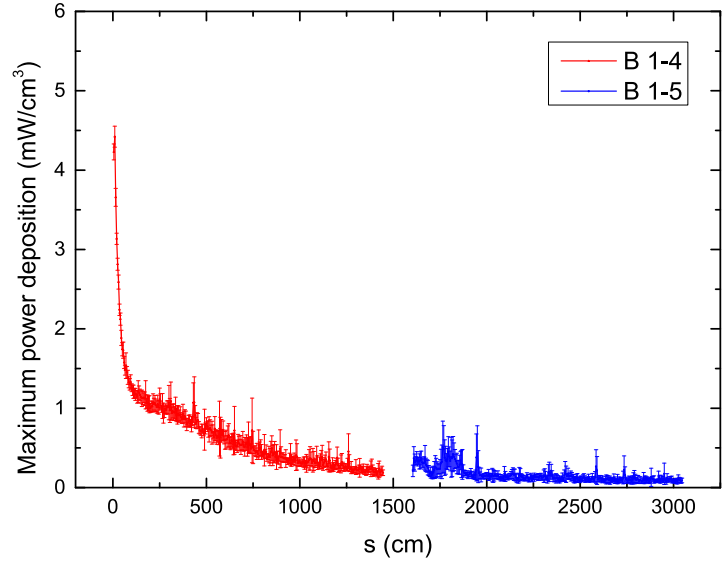

FIG. 17.  Maximum power deposition along the two most exposed cold dipoles, which are closest to the first protective collimator

The maximum power deposition is 4.5 mW/cm$^3$, which is below the quench limit 5-10 mW/cm$^3$. At present, additional shielding following the protective collimator for the SC coils is not necessary. However, this issue needs to be re-considered for the shorter beam lifetime or possible upgrading plane, some related results have been carried out in FCC-hh [55].

## V. CONCLUSIONS

A novel collimation method for future proton-proton colliders is proposed, which arranges both transverse and momentum collimation systems in the same cleaning insertion and employs SC quadrupoles in the transverse collimation section. The design and simulation results with the SPPC parameters show the convincing effectiveness of the method. Two major features are: the momentum collimation section just following the transverse collimation section can effectively clean the particles with large momentum deviation produced in the transverse collimators thus practically eliminate beam loss in the downstream DS section; the application of SC quadrupoles in the transverse collimation section can help create one more collimation stage which turns out to be very effective in reducing beam loss in the momentum collimation and experimental sections. Simulations with the FLUKA code have proven that with some protection design, the SC magnets in the collimation section can be safe from quenches caused by the radiation effect. The main design goal of collimation inefficiency $3.55\times10^{-7}$ m$^{-1}$ at the cold regions can be fulfilled very well. Although the details have been carried with the SPPC parameters, the method should be more general for proton colliders of such scale.

Eliminating the great risk of particle loss in the cold region due to cleaning beam halo particles, for the colliders with ultra-high luminosity, it is foreseen that the collision debris gives significant contribution of particle losses around the experimental points. With the great challenges to the optical design and protection scheme of experiment insertions, it may be an effective to apply the momentum collimation method in the same long straight sections to avoid the cold losses in the downstream DSs of the experiment regions. This work should be done in the future.

## Acknowledgments

The authors thank James Molson of CERN and Jianshu Hong of IHEP for supporting the MERLIN code at IHEP, and also the SPPC colleagues for discussions. This work was supported by the National Natural Science Foundation of China (Projects: 11575214, 11527811, 11235012).

## APPENDIX: BRIEF OF THE DESIGN FEATURES AND MAIN PARAMETERS OF THE SPPC

SPPC (Super Proton-Proton Collider) is a next-generation proton-proton collider aiming for energy-frontier physics, especially for beyond-Standard Model research. It is the second phase of the CEPC-SPPC project which was proposed by Chinese scientists, and the two colliders use the same tunnel [10]. As a future proton-proton collider, SPPC will collide protons at a center of mass energy of 75 TeV, with circumference of 100 km, a nominal luminosity of $1.01 \times 10^{35}$ cm$^{-2}$s$^{-1}$ per IP. There is also an upgrading plan with higher energy (125-150 TeV). In baseline design, full iron-based HTS technology will be used in the SC magnets, and the field strength of the main dipoles is 12 T. Figure 18 shows the layout of the SPPC and Table IV shows the main parameters given by the CEPC Conceptual Design Report (CDR) [39].

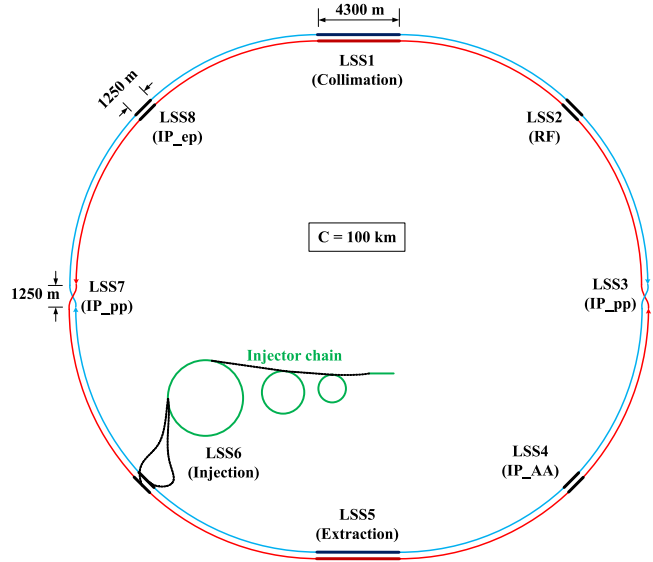

FIG. 18. The layout of SPPC

TABLE IV. Main parameters of SPPC

| Parameter | Unit | CDR |
|---|---|---|
| Circumference | km | 100 |
| C.M. energy | TeV | 75 |
| Dipole field | T | 12 |
| Injection energy | TeV | 2.1 |
| Number of IPs | - | 2 |
| Nominal luminosity per IP | $cm^{-2}s^{-1}$ | $1.0\times10^{35}$ |
| Beta function at collision | m | 0.75 |
| Circulating beam current | A | 0.7 |
| Bunch separation | ns | 25 |
| Bunch population | - | $1.5\times10^{11}$ |
| Number of bunches | - | 10080 |
| Normalized emittance | μm | 2.4 |
| SR power per beam | MW | 1.1 |
| SR heat load per aperture @arc | W/m | 13 |

SPPC is a complex accelerator facility and will be able to support research in different fields of physics, similar to the multi-use accelerator complex at CERN. Besides the energy frontier physics program in the collider, the beams from each of the four accelerators in the injector chain can also support their own physics programs. The four stages, shown in Fig. 19, are a proton linac (p-Linac), a rapid cycling synchrotron (p-RCS), a medium-stage synchrotron (MSS) and the final stage synchrotron (SS). This research can occur during periods when beam is not required by the next-stage accelerator.